\documentclass[3p,times,twocolumn]{elsarticle}

\usepackage{ecrc}
\usepackage{amsmath}

\volume{00}

\firstpage{1}

\journalname{Nuclear Physics B Proceedings Supplement}

\runauth{D. Boito, M. Jamin, R. Miravitllas}

\jid{nppp}

\jnltitlelogo{Nuclear Physics B Proceedings Supplement}

\usepackage{amssymb}

\usepackage[figuresright]{rotating}

\newcommand{\MSbar}{{\overline{\rm MS}}}
\newcommand{\sfrac}[2]{\mbox{$\frac{#1}{#2}$}}
\newcommand{\ah}{\hat a}
\newcommand{\cO}{{\cal O}}

\newcommand{\eqn}[1]{(\ref{#1})}
\newcommand{\eq}[1]{Eq.~(\ref{#1})}

\def\alphahat{\hat \alpha_s}

\def\beq{\begin{equation}}
\def\eeq{\end{equation}}
\def\bea{\begin{eqnarray}}
\def\eea{\end{eqnarray}}

\def\nn{\nonumber}

\begin{document}

\begin{frontmatter}



\title{Scheme variations of the QCD coupling and tau decays\tnoteref{Conf}}
 \tnotetext[Conf]{Talk given at ``The 14th International Workshop on Tau Lepton Physics'',
19--23 September 2016, IHEP, Beijing, China}
 
\author[label1]{Diogo Boito\corref{cor1}}
\cortext[cor1]{Speaker}
\ead{boito@ifsc.usp.br}

\author[label2,label3]{Matthias Jamin}
 
\author[label2]{Ramon Miravitllas}

 \address[label1]{Instituto de F\'isica de S\~ao Carlos, Universidade de S\~ao Paulo,
CP 369, 13560-970, S\~ao Carlos SP, Brazil}
  
\address[label2]{IFAE, BIST, Campus UAB, 08193 Bellaterra (Barcelona) Spain}

\address[label3]{ICREA, Pg.~Llu\'\i s Companys 23, 08010 Barcelona, Spain}




\begin{abstract}
The QCD coupling, $\alpha_s$, is not a physical observable since it
depends on conventions related to the renormalization procedure.  Here
we discuss a redefinition of the coupling where changes of scheme are parametrised by a single
parameter $C$. The new coupling is denoted $\hat
\alpha_s$ and its running is scheme
independent. Moreover, scheme variations become completely analogous
to renormalization scale variations. We discuss how the coupling $\hat \alpha_s$ can be used in order to optimize predictions for the inclusive hadronic
decays of the tau lepton. Preliminary investigations of the $C$-scheme in the presence of higher-order terms of the perturbative series are discussed here for the first time. 
\end{abstract}


\end{frontmatter}


\section{Introduction}
\label{}

The perturbative expansion in the strong coupling $\alpha_s$ is the
main approach to predictions in quantum chromodynamics (QCD) at
sufficiently high energies. However, the expansion parameter,
$\alpha_s$, is not a physical observable of the theory. Its definition
 carries a dependence on conventions related to the
renormalization procedure, such as the renormalization scale and
renormalization scheme.  Physical observables should, of course, be
independent of any such conventions. This requirement leads, in the
case of the renormalization scale, to well defined Renormalization
Group Equations (RGE) that must be satisfied by physical quantities.
The situation regarding the renormalization scheme is more complicated and
perturbative computations are, most often, performed in conventional schemes
such as $\MSbar$~\cite{bbdm78}.

In this work we discuss a new definition of the QCD coupling, that we
denote $\alphahat$, recently introduced in Ref.~\cite{BJM16},
and its applications to the QCD description of inclusive hadronic
$\tau$ decays. The running of this new coupling is renormalization
scheme independent, i.e. in its $\beta$ function only scheme
independent coefficients intervene. The scheme dependence of $\hat
\alpha_s$ is parametrised by a single continuous parameter $C$.  The
evolution of $\hat \alpha_s$ with respect to this new parameter is
governed by the same $\beta$ function that governs the scale
evolution.  We  refer to the coupling $\hat \alpha_s$ as the $C$-scheme coupling.

An important aspect is the fact that perturbative expansions in
$\alpha_s$ are divergent series that are assumed to be asymptotic
expansions to a ``true'' value, which is
unknown~\cite{Renormalons}.\footnote{F. Dyson formulated the first form
  of this reasoning in 1952, in the context of Quantum
  Electrodynamics~\cite{FD52}.} In this spirit, different schemes
correspond to different asymptotic expansions to the same scheme
invariant physical quantity, and should be interpreted as such. One can then
use the parameter $C$ to interpolate between perturbative series with
larger or smaller coupling values, and exploit this dependence in
order to optimize the predictions for observables of the theory.

The idea of exploiting the scheme dependence in order to optimize the
series differs from the approach of other celebrated methods used for
the optimisation of perturbative predictions. In methods such as
Brodsky-Lepage-Mackenzie (BLM)~\cite{Brodsky:1982gc} or
the Principle of Maximum
Conformality~\cite{Brodsky:2012rj,Mojaza:2012mf} the idea is to obtain a scheme
independent result through a well defined algorithm for setting the renormalization scale, regardless of the intermediate scheme used for the
perturbative calculation (which most often is $\MSbar$). The
``effective charge" method~\cite{Grunberg:1982fw}, on the other hand,
involves a process dependent definition of the coupling. In the
procedure described here, one defines a process independent class of
schemes, parametrised by the parameter $C$.  The optimal value of $C$
must be set independently for each process considered.

We begin with the scale running of the QCD coupling that
we write as 
\begin{equation}
\label{bfun}
-\,Q\,\frac{{\rm d}a_Q}{{\rm d}Q} \,\equiv\, \beta(a_Q) \,=\,
\beta_1\,a_Q^2 + \beta_2\,a_Q^3 + \beta_3\,a_Q^4 + \cdots 
\end{equation}
We will work with $a_Q \equiv \alpha_s(Q)/\pi$, with $Q$ being a
physically relevant scale. Since the recent five-loop computation of
Ref.~\cite{bck16}, the first five coefficients of the QCD
$\beta$-function are known analytically.  The coefficients $\beta_1$
and $\beta_2$ are scheme independent.

Let us consider a scheme
transformation to a new coupling $a'$, which, perturbatively, takes the general form
\begin{equation}
\label{ap}
a' \,\equiv\, a + c_1\,a^2 + c_2\,a^3 + c_3\,a^4 + \cdots \,
\end{equation}
The QCD scale $\Lambda$ is also different in the two schemes and
obeys the relation
\begin{equation}
\label{Lambdap}
\Lambda' \,=\, \Lambda\,{\rm e}^{c_1/\beta_1}.
\end{equation}
The shift in $\Lambda'$ depends only on a single constant~\cite{cg79}, governed by  $c_1$ of \eq{ap}.
This fact motivates the definition of the new coupling $\ah_Q$, which 
is scheme invariant except for shifts in $\Lambda$ parametrised by a 
parameter $C$ as
\begin{eqnarray}
\label{ahat}
\frac{1}{\hat a_Q} + \frac{\beta_2}{\beta_1} \ln\hat a_Q \,&\equiv&\,
\beta_1 \Big( \ln\frac{Q}{\Lambda} + \frac{C}{2} \Big) \nn \\
&& \hspace{-18mm} \,=\, \frac{1}{a_Q} + \frac{\beta_1}{2}\,C +
\frac{\beta_2}{\beta_1}\ln a_Q - \beta_1 \!\int\limits_0^{a_Q}\,
\frac{{\rm d}a}{\tilde\beta(a)},
\end{eqnarray}
where
\begin{equation}
\frac{1}{\tilde\beta(a)} \,\equiv\, \frac{1}{\beta(a)} - \frac{1}{\beta_1 a^2}
+ \frac{\beta_2}{\beta_1^2 a}
\end{equation}
is free of singularities in the limit $a\to 0$ and  we have used the scale invariant form of $\Lambda$. The coupling $\ah_Q$ is a function of the parameter $C$ but we do not make this  dependence explicit  to keep the notation simple.
The definition of \eq{ahat} should be interpreted in perturbation theory in an
iterative sense, which allows one to deduce the corresponding
coefficients $c_i$ of \eq{ap} (their explicit expressions are given
 in \cite{BJM16} using the $\MSbar$ as the input scheme). One should remark that a combination
similar to~\eqn{ahat}, but without the logarithmic term on the left-hand side,
was already discussed in Refs.~\cite{byz92,ben93}. However, without this term, 
an unwelcome logarithm of $a_Q$ remains in the perturbative relation between
the couplings $\ah_Q$ and $a_Q$. This non-analytic term is avoided by the
construction of Eq.~\eqn{ahat}. 

\begin{figure}
\includegraphics[width=0.47\textwidth]{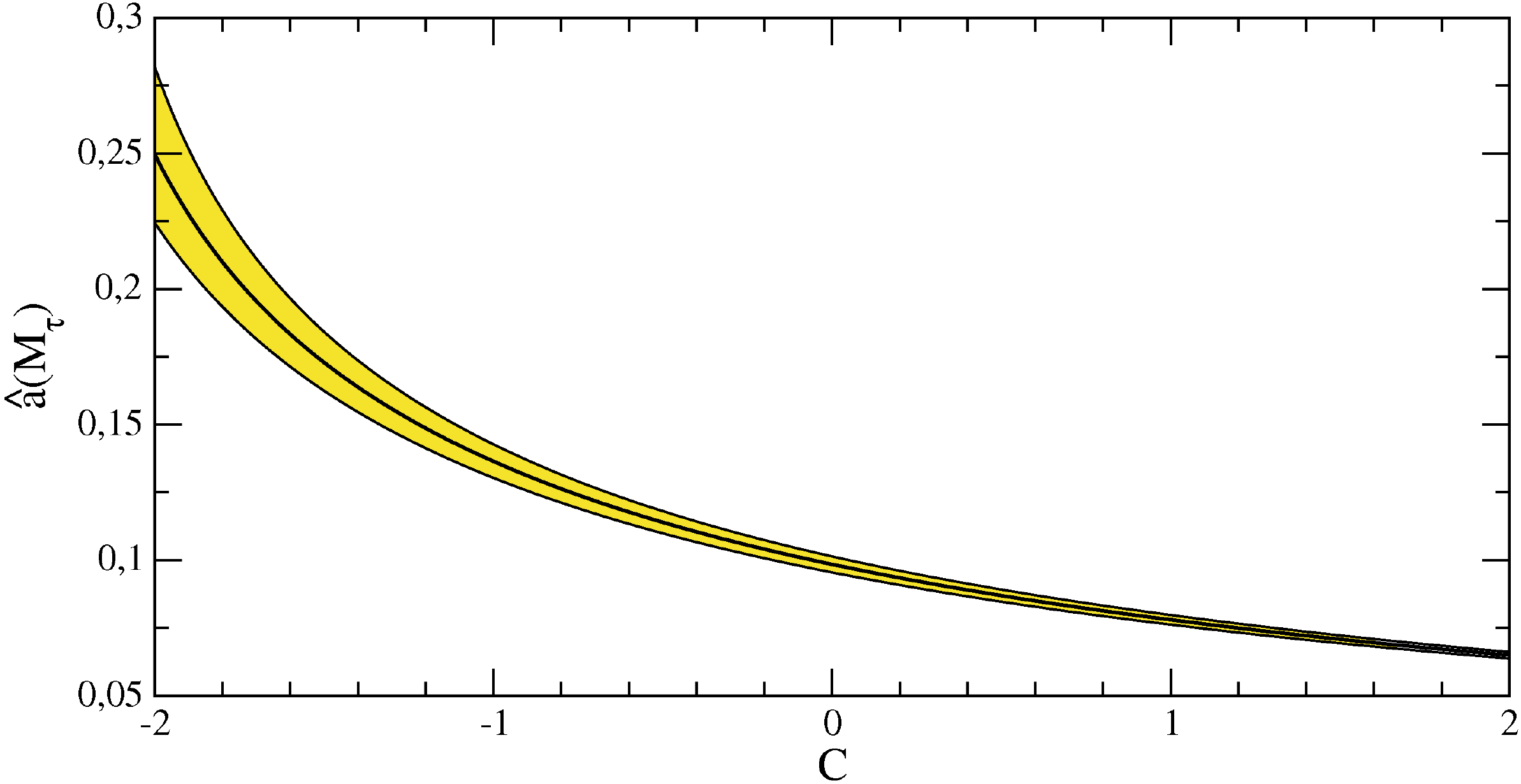}
\caption{The coupling $\ah(M_\tau)$ according to Eq.~\eqn{ahat} as a function
of $C$, and for the $\MSbar$ input value $\alpha_s(M_\tau)=0.316(10)$. The
yellow band corresponds to the $\alpha_s$ uncertainty.\label{fig1}}
\end{figure}

From the definition of the new coupling $\ah_Q$ we can derive its $\beta$ function that reads
\beq
\label{betahat}
-\,Q\,\frac{{\rm d}\ah_Q}{{\rm d}Q} \,\equiv\, \hat\beta(\ah_Q) \,=\,
\frac{\beta_1 \ah_Q^2}{\left(1 - \sfrac{\beta_2}{\beta_1}\, \ah_Q\right)} .
\eeq 
The function $\hat \beta$ takes a simple form and is scheme independent since only the coefficients $\beta_1$ and $\beta_2$ intervene. The evolution with 
 the parameter $C$ obeys an analogous  equation
\beq
-2 \frac{{\rm d}\ah_Q}{{\rm d}C} =\,
\frac{\beta_1 \ah_Q^2}{\left(1 - \sfrac{\beta_2}{\beta_1}\, \ah_Q\right)}.
\eeq
Therefore, there is a complete analogy between the coupling evolution
with respect to the scale and with respect to the scheme parameter
$C$. The dependence of $\ah_Q$ on $C$ is displayed in Fig.~\ref{fig1}
 using the $\MSbar$ as  input scheme and setting the scale to the $\tau$ mass, $M_\tau$. The new
coupling becomes smaller for larger values of $C$ and perturbativity breaks down
for  values below roughly $C=-2$. Therefore, we restrict our analysis to 
 $C\geq -2$.

\section{\boldmath Application to $\tau$ decays}

As a phenomenological application of the $C$-scheme coupling, we focus here on the
perturbative expansion of the total $\tau$ hadronic width. The chief
observable is the ratio $R_\tau$ of the total hadronic branching
fraction to the electron branching fraction. It is conventionally decomposed
as
\begin{equation}
R_\tau \,=\, 3\, S_{\rm EW} (|V_{ud}|^2 + |V_{us}|^2)\, ( 1 + \delta^{(0)}
+ \cdots),
\end{equation}
where $S_{\rm EW}$ is an electroweak correction and $V_{ud}$, as well
as $V_{us}$, CKM matrix elements. Perturbative QCD is encoded in
$\delta^{(0)}$ (see Refs.~\cite{bnp92,bj08} for details) and the
ellipsis indicate further small sub-leading corrections.  The
calculation of $\delta^{(0)}$ is performed from a contour integral of the so 
called Adler function in
the complex energy plane, exploiting analyticity properties, which
allows one to avoid the low energy region where perturbative QCD is
not valid. In doing so, one must adopt a procedure in order to deal
with the renormalization scale. The scale logarithms can be summed
either before or after performing the contour integration.  The first
choice, where the integrals are performed over the running QCD
coupling, is called Contour Improved Perturbation Theory (CIPT), while
the second, where the coupling is evaluated at a fixed scale and the
integrals are performed over the logarithms, is called Fixed Order
Perturbation Theory (FOPT). 

Analytic results for the coefficients of the
Adler function are available up to five loops, or $\alpha_s^4$ \cite{bck08}.
Here we consider an estimate for the yet unknown fifth order coefficient
of the Adler function,  namely  $c_{51}=283$ \cite{bj08}.

In FOPT, the perturbative series of $\delta^{(0)}(a_Q)$ in terms of the $\MSbar$
coupling $a_Q$ is given by \cite{bck08,bj08}
\begin{align}
\label{del0}
\delta_{\rm FO}^{(0)}(a_Q) =
a_Q + 5.202a_Q^2 + 26.37a_Q^3 + 127.1a_Q^4 +\cdots
\end{align}
In the $C$-scheme coupling $\ah_Q$, the expansion for
$\delta_{\rm FO}^{(0)}$ is
\begin{align}
\label{del0ah}
&\delta_{\rm FO}^{(0)}(\ah_Q) = \ah_Q + (5.202 + 2.25 C)\,\ah_Q^2 \nn \\
& \hspace{0.2cm} + (27.68 + 27.41 C + 5.063 C^2)\,\ah_Q^3 \nn\\
& \hspace{0.2cm} + (148.4 + 235.5 C + 101.5 C^2 + 11.39 C^3)\,\ah_Q^4\nn\\ 
&  \hspace{0.2cm} + \cdots 
\end{align}

\begin{figure}
\includegraphics[height=3.7cm]{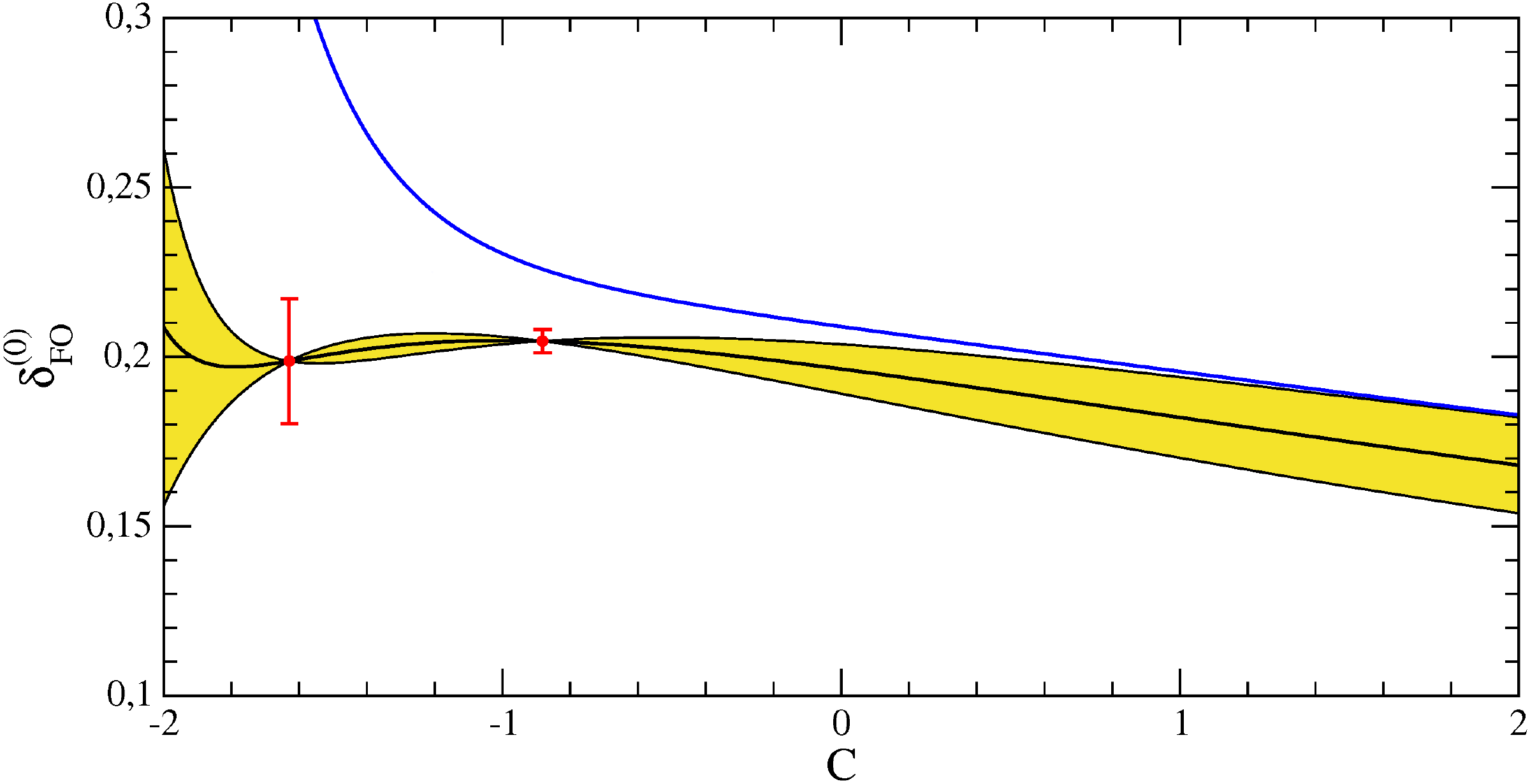}
\caption{$\delta_{\rm FO}^{(0)}(\ah_Q)$ of Eq.~\eqn{del0ah} as a function of
$C$. The yellow band arises from either removing or doubling the fifth-order
term. In the red dots, the $\cO(\ah^5)$ vanishes, and $\cO(\ah^4)$ is taken as
the uncertainty. 
For further explanation, see the text.
\label{fig3}}
\end{figure}

{\noindent In Fig.~\ref{fig3}, we display $\delta_{\rm FO}^{(0)}(\ah_Q)$ as a function
of $C$. Assuming $c_{5,1}=283$, the yellow band  corresponds to removing
or doubling the $\cO(\ah^5)$ term. A  plateau is
found for $C\approx -1$. Taking $c_{5,1}=566$ and then doubling the $\cO(\ah^5)$
results in the blue curve that does not show this stability. Hence, this scenario
 is disfavoured. In the red dots, which lie at $C=-0.882$ and $C=-1.629$,
the $\cO(\ah^5)$ correction vanishes, and the $\cO(\ah^4)$ term is taken as
the uncertainty, in the spirit of asymptotic series. The point to the right has a substantially smaller error, and yields}
\begin{equation}
\label{del0oa5zero}
\delta_{\rm FO}^{(0)}(\ah_{M_\tau},C=-0.882) \,=\,
0.2047 \pm 0.0034 \pm 0.0133 \,.
\end{equation}
The second error covers the uncertainty of $\alpha_s(M_\tau)$.
In this case, the direct $\MSbar$ prediction of Eq.~\eqn{del0} is 
\begin{equation}
\label{del0MSb}
\delta_{\rm FO}^{(0)}(a_{M_\tau}) \,=\, 0.1991 \pm 0.0061 \pm 0.0119\,\,\,\,\, (\MSbar) \,.
\end{equation}
This value is somewhat lower, but within $1\,\sigma$ of the higher-order
uncertainty. 

In CIPT, contour integrals over the running coupling
have to be computed, and hence the result cannot be given in analytical form.
The general behaviour is very similar to FOPT,
with the exception that now also for $c_{5,1}=566$ a zero of the $\cO(\ah^5)$
term is found. Employing the value of $C$ which leads to the smaller uncertainty one finds
\begin{equation}
\label{del0CIoa5z}
\delta_{\rm CI}^{(0)}(\ah_{M_\tau},C=-1.246) \,=\,
0.1840 \pm 0.0062 \pm 0.0084 \,.
\end{equation}
As has been discussed many times in the past (see e.g.~\cite{bj08})
the CIPT prediction lies substantially below the FOPT results. On the
other hand, the parametric $\alpha_s$ uncertainty in CIPT turns out to
be smaller.

\section{Higher-order terms}

The behaviour of the series at higher orders is not known
exactly. However, realistic models of the Adler function can be
constructed in the Borel plane, in which the singularities of the
function, namely its renormalon content, is partially
known~\cite{Renormalons}.  In Ref.~\cite{bj08} (see also
Ref.~\cite{BBJ14}), models of the Adler function were constructed
using the leading renormalons, that largely dominate the higher-order
behaviour of the perturbative series.  The model is matched to the
exactly known coefficients in order to fully reproduce QCD for terms
up to $a_Q^4$.  This allows for a complete reconstruction of the
series, to arbitrarily high orders in the coupling, and, moreover, one
is able to obtain the ``true'' value of the asymptotic series by means
of the Borel sum. In fact, the series is not strictly Borel summable
because infra-red renormalons obstruct integration on the positive
real axis. The ``true'' value has, therefore, an inherent ambiguity
that stems from the prescription adopted to circumvent the
singularities along the contour of integration. This ambiguity is
related to non-perturbative physics~\cite{Renormalons, bj08}.

\begin{figure}
\includegraphics[width=1\columnwidth,angle=0]{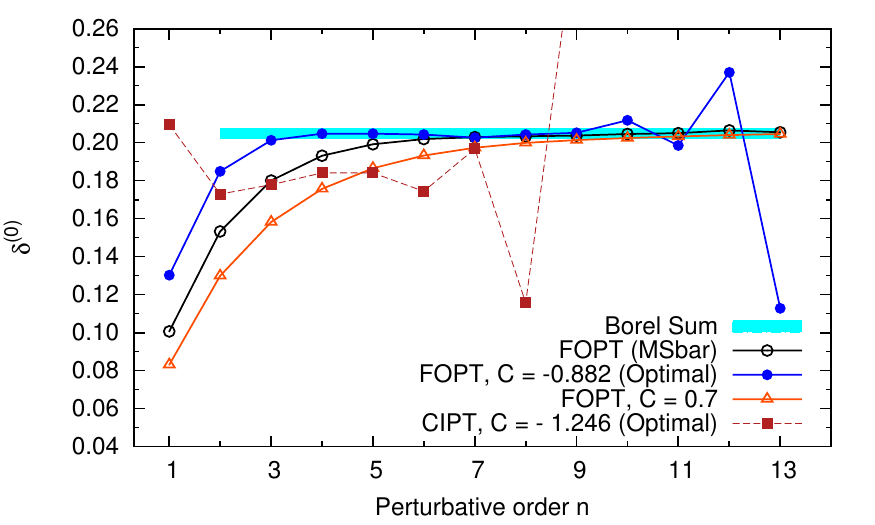}\label{Delta0}
\caption{Four series for $\delta^{(0)}$ with higher-order coefficients from the central model of Ref.~\cite{bj08}. In all cases $\alpha_s(M_\tau)=0.316$ which corresponds to the central  value of the present world average~\cite{PDG16}. The optimised FOPT (filled circles) and CIPT (filled squares) series can be compared with the FOPT $\MSbar$ results (empty circles) and FOPT for $C=0.7$ (triangles). The shaded band gives the Borel summed result, the ``true" value of the series, with its associated ambiguity~\cite{BJM16}.  }\vspace{-0.5cm} \end{figure}

Here we perform a preliminary investigation of the behaviour of
$\delta^{(0)}$ at higher orders using the $C$-scheme coupling.
The Adler function coefficients for terms higher than $a_Q^5$ 
are obtained in the $\MSbar$ scheme from the central model of Ref.~\cite{bj08}.
The series can then be translated to the $C$-scheme by means of the perturbative
relation between the couplings $a_Q$ and $\hat a_Q$~\cite{BJM16}.
  Fig.~\ref{Delta0} shows
four different series that should approach the same Borel summed
result, showed as a horizontal band. The four series use as input the
coefficients exactly known in QCD with the addition of the estimate
$c_{5,1}=283$. One observes that the optimised version of $\delta_{\rm
  FO}^{(0)}$ (filled circles) approaches the Borel sum of the series
faster than the $\MSbar$ result (empty circles). Of course, because
the optimised series has a larger coupling (see Fig.~1) asymptoticity
sets in earlier and the divergent character  is clearly
visible already around the 10th order.  The FOPT result with $C=0.7$
shows that smaller couplings do not necessarily lead to a better
approximation at lower orders, requiring many more terms to give a
good approximation to the Borel summed result. Finally, the optimal
CIPT series does not give a good approximation to the Borel summed
result (this is also the case in the
$\MSbar$~\cite{bj08}). Unfortunately, the use of the $C$-scheme
coupling does not make the CIPT prediction closer to FOPT. The
$C$-scheme FOPT, on the other hand, is in excellent agreement with the
central Borel model which suggests that  FOPT should be the
favoured expansion.

\vspace{-0.5cm}
\section*{Acknowledgements}
\vspace{-0.3cm}
It is a pleasure to thank the organisers of this very fruitful
meeting.  DB is supported by the S\~ao Paulo Research Foundation
(FAPESP) grant 2015/20689-9, and by CNPq grant 305431/2015-3.  The
work of MJ and RM has been supported in part by MINECO Grant number
CICYT-FEDER-FPA2014-55613-P, by the Severo Ochoa excellence program of
MINECO, Grant SO-2012-0234, and Secretaria d'Universitats i Recerca
del Departament d'Economia i Coneixement de la Generalitat de
Catalunya under Grant 2014 SGR~1450.

\bibliographystyle{elsarticle-num}
\bibliography{ShortBib}







\end{document}